\documentclass[sigconf, 10pt, nonacm]{acmart}
\usepackage{xspace}
\usepackage{graphicx} 
\usepackage{subcaption}

\usepackage[compact]{titlesec}
\renewcommand{\paragraph}[1]{\vspace{0.02in}\noindent {\bf #1}~}

\usepackage{tabularray}
\SetTblrInner{rowsep=0pt}
\UseTblrLibrary{amsmath}

\usepackage{caption}
\newcommand{\eg}{{\it e.g.}\xspace}

\newcommand{\etc}{{\it etc.}\xspace}

\newcommand{\cloud}{WAN-transit\xspace}
\newcommand{\wan}{\cloud}
\newcommand{\inet}{Internet-transit\xspace}

\newcommand{\rev}[2]{}

\newcommand{\optimal}{\texttt{MinLatency}\xspace}
\newcommand{\baseline}{\texttt{Baseline}\xspace}
\newcommand{\asprep}{\texttt{ASPrep}\xspace}

\title{Latency-Aware Inter-domain Routing}

\author{Shihan Lin}
\affiliation{%
    \institution{Duke University}
    \city{Durham}
    \country{USA}}

\author{Yi Zhou}
\affiliation{%
    \institution{Duke University}
    \city{Durham}
    \country{USA}}

\author{Xiao Zhang}
\authornote{Xiao Zhang was with Duke University at the time this work was conducted. He is now
    with Cisco ThousandEyes.}
\affiliation{%
    \institution{Cisco ThousandEyes}
    \city{Raleigh}
    \country{USA}}

\author{Todd Arnold}
\authornote{The views expressed herein are those of the authors and do not reflect the position of the US Military Academy, Department of the Army, or Department of Defense.}
\affiliation{%
    \institution{U.S. Military Academy}
    \city{West Point}
    \country{USA}}

\author{Ramesh Govindan}
\affiliation{%
    \institution{University of Southern California}
    \city{Los Angeles}
    \country{USA}}

\author{Xiaowei Yang}
\affiliation{%
    \institution{Duke University}
    \city{Durham}
    \country{USA}}

\makeatletter
\let\@authorsaddresses\@empty
\makeatother

\begin{abstract}
  Despite efforts from cloud and content providers to lower latency to acceptable levels for current and future services (\eg, augmented reality or cloud gaming), there are still opportunities for improvement. 
  A major reason that traffic engineering efforts are challenged to lower latency is that the Internet's inter-domain routing protocol, the Border Gateway Protocol, is oblivious to any performance metric, and circuitous routing is still pervasive.

  In this work, we propose two implementation modifications that networks can leverage to make BGP latency-aware and reduce excessive latency inflation.  These proposals, latency-proportional AS prepending and local preference neutralization, show promise towards providing a method for propagating abstract latency information with a reasonable increase in routing overhead.
\end{abstract}

\usepackage{cleveref}
\crefname{section}{\S}{\S\S}
\Crefname{section}{Section}{Sections}
\crefname{figure}{Figure}{Figure}
\Crefname{figure}{Figure}{Figures}

\usepackage[nolist]{acronym}
\begin{acronym}
  \acro{rtt}[RTT]{Round Trip Time}
  \acro{bgp}[BGP]{Border Gateway Protocol}
  \acro{as}[AS]{Autonomous System}
  \acrodefplural{as}[ASes]{Autonomous Systems}
\end{acronym}

\begin{document}

\maketitle
\pagestyle{plain}

\section{Introduction}
\label{sec:intro}



The Border Gateway Protocol (BGP)~\cite{rfc4271_bgp} has long been known to
sometimes find excessively long routes, despite ISP efforts to ensure shortest paths for their customers~\cite{SIGCOMM03-PathInflation}. More recently, researchers have also shown that IP anycast systems, including both root DNS servers and CDNs, may
experience anycast path inflation, where clients are routed to anycast
replicas continents away, due to BGP's poor path choices~\cite{Li2018Internet,iLiu2007Two,JH2015Analysing,de-GDNS-19TNS,SIGCOMM23-RegionalCDN,rizvi2024anycast,twotale-li-sigcomm21}. Cloud providers, too, have encountered this problem~\cite{beatbgp-todd-hotnet18,INFOCOM20-PrivateWAN}.

A direct consequence of path inflation is latency inflation.
End-to-end \acp{rtt} can be inflated by tens of
milliseconds if traffic is routed along sub-optimal paths. 
Applications that require low latency~\cite{hotnets-edge-computing} -- such as cloud gaming, web, video
conferencing, and online streaming -- may be unusable in these cases.
To address path inflation, cloud providers have employed software-defined WANs to make latency-aware choices for outbound traffic~\cite{schlinker2017edge,espresso}; this does not ensure latency-aware \textit{inbound} traffic routing.
A line of prior work~\cite{Chen2015End, detour,Arrow,resilient-andersen-2001} has explored routing on a latency-aware overlay, but this requires significant deployment of new overlay infrastructure.

On the flip side, structural changes to the Internet, such as flattening, appear to have ameliorated the problem somewhat. Recent work~\cite{beatbgp-todd-hotnet18} has shown that BGP finds low latency and high throughput paths most of the time, so the case for replacing BGP~\cite{trotsky-sigcomm19,raja2017dbgp,nira,chuat2022complete} has become weaker.




However, adding latency awareness to BGP is still highly desirable for services with a large footprint where higher latency adversely impacts revenue. 
For example, prior work found that 9\% of vantage points fail to adopt a path with lower latency to
connect to Google's US-Central datacenter~\cite{INFOCOM20-PrivateWAN}.
We focus on two such efforts for lowering latency---the use of private WANs by cloud providers to ensure low-latency access to cloud customers and the use of IP anycast by CDNs for delivering low-latency content---and examine case studies for improvement (\cref{sec:motivation}).






For these settings, we posit that a relatively small, incrementally deployable change to BGP \textit{implementations}--leaving the protocol unmodified, together with an incentive-aligned change to business agreements might help address the tail latency inflation in today's Internet. 
Specifically, we propose \textit{latency-proportional AS path prepending} (\cref{subsec:ASPrep}) in which an AS prepends its own AS number (ASN) $n$ times, where $n$ is proportional to the propagation latency that would be incurred traversing its own infrastructure for received traffic. 
We also propose \textit{local preference neutralization}~(\cref{subsec:localpref}), whereby a cloud or content provider covers both the ``uphill'' and ``downhill'' transit costs for its traffic~\cite{gao2021policy}, enabling other ASes to make latency-aware choices by making their BGP best path selection fallback to AS path length.
Cloud and content providers have an incentive to do this when the latency-sensitive service revenue offsets the transit costs.



To gain insights into how well the proposed techniques work, we use a
router-level Internet-topology from CAIDA~\cite{ITDK}
to simulate the above BGP modifications.
We compare its latency improvement with the current BGP
and a protocol that achieves theoretical min-latency~(\S~\ref{sec:simulation}).
Our simulation results show that compared to vanilla
BGP, our proposed implementation modifications of introducing latency-awareness into BGP 
reduce the 90th percentile of end-to-end latency 
by more than 31\% and incur about half of the message overhead
compared to the theoretical min-latency routing protocol.

\paragraph{Ethical concerns:} This work does not raise any ethical concerns.

\section{BGP Path Inflation}
\label{sec:motivation}

In this section, we first summarize BGP's best path selection
algorithm. We then use three concrete examples to analyze the root
causes of BGP path inflation.

\begin{figure*}[t]
    \begin{minipage}[c][][b]{0.35\linewidth}
        \centering
        \includegraphics[width=\linewidth,trim={10pt 10pt 10pt 10pt}]{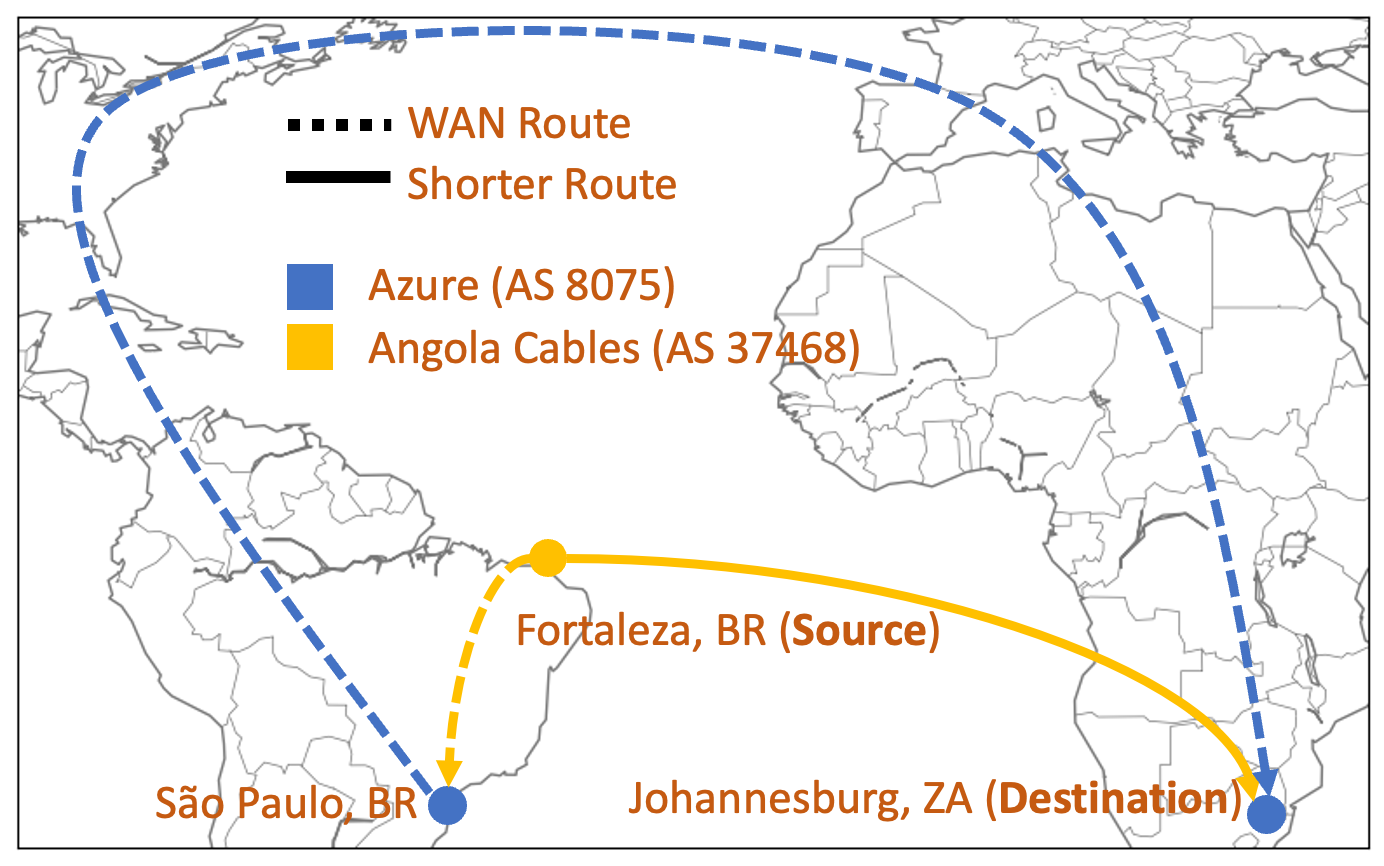}
        \subcaption{Case~1\label{fig:case1}}
    \end{minipage}
    \begin{minipage}[c][][b]{0.31\linewidth}
        \centering
        \includegraphics[width=\linewidth]{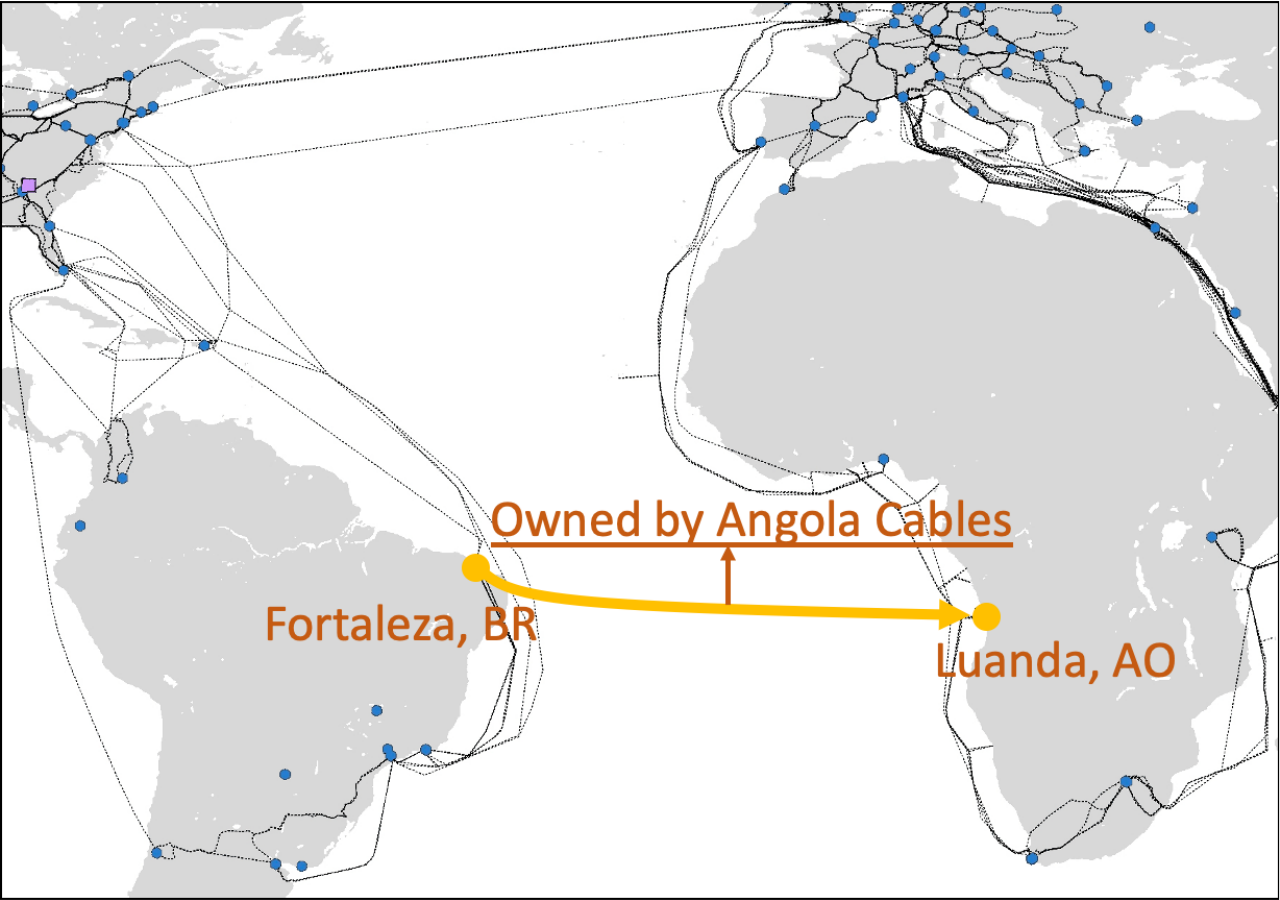}
        \subcaption{Network map for Case~1\label{fig:network_map}}
    \end{minipage}
    \begin{minipage}[c][][b]{0.324\linewidth}
        \centering
        \includegraphics[width=\linewidth,trim={10pt 10pt 10pt 10pt}]{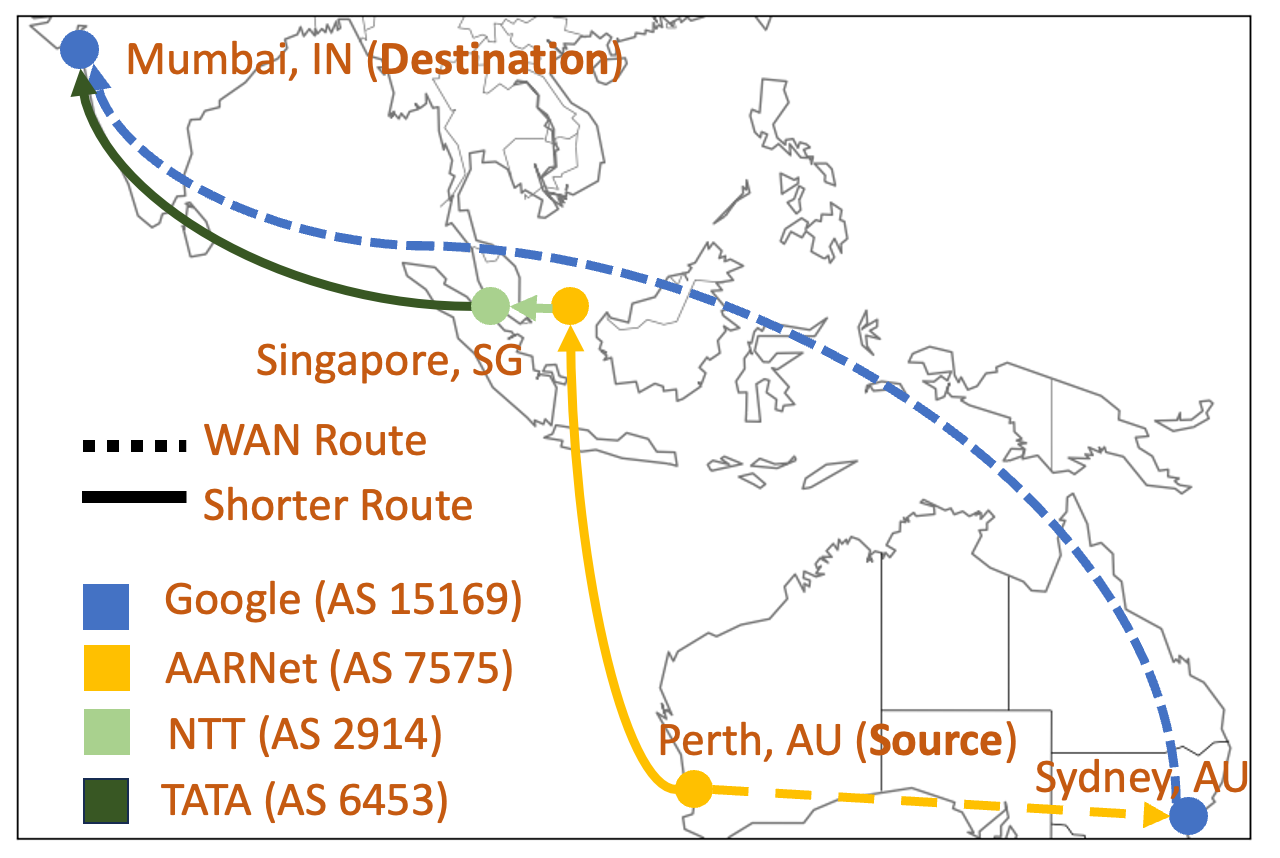}
        \subcaption{Case~2\label{fig:case2}}
    \end{minipage}
    \caption{\footnotesize\bfseries (a)~A RIPE Atlas probe located in Fortaleza, BR takes the \wan (dashed line) and the \inet (solid line) service to reach an Azure VM located in Johannesburg, ZA. Colors encode path segments in different ASes. (b)~Part of network infrastructure map of Azure~\cite{MicrosoftMap} (black lines) and Angola Cables~\cite{AngolaMap} (yellow line). (c)~A RIPE Atlas probe located in Perth, AU takes the \wan (dashed line) and the \inet (solid line) to reach a Google VM located in Mumbai, IN.}
\end{figure*}

\subsection{BGP Best Path Selection}
\label{sec:bgp_select}


Unlike intra-domain routing protocols~\cite{RFC2328-OSPF,RFC1142-ISIS}, BGP~\cite{rfc4271_bgp} does not use a single optimization metric for route selection. 
BGP's best path selection process compares route attributes in a pre-determined order, starting with the local preference attribute, which captures the policy preferences of an AS.\textbf{}
For example, ASes often prefer customer-learned routes over peer-learned and prefer peer-learned over provider routes~\cite{ToN01-GaoRexford}. For the routes to the same destination with the same local preference, BGP routers use a series of tie-breaks to pick the best route. In order, these are: AS path length, route origin, the Multi-Exit Discriminator (MED), an eBGP/iBGP, interior routing costs, \etc.

\subsection{Examples of Path Inflation in BGP}
\label{sec:example}
\rev{C}{Adding evidence that this is a common problem and latency variations are not due to congestion.}

BGP policy choices, as well as Internet connectivity, cabling, and peering contribute to the Internet path inflation~\cite{SIGCOMM03-PathInflation}. Investing in more direct paths~\cite{NSDI22-cisp} and ensuring broader
peering can reduce path and latency inflation, but today's Internet still exhibits instances of latency inflation.
We use examples from real-world network traces to illustrate high latency inflation instances: in these examples, shorter valley-free paths~\cite{ToN01-ASRel} exist, 
yet BGP fails to find those paths.

\paragraph{Case 1:} The cloud provider Azure
peers with the ISP Angola Cables in both S\~{a}o Paulo, Brazil
and Johannesburg, South Africa.
Azure has a datacenter in
Johannesburg~\cite{azure-pops}, owns a private WAN, and offers its customer's VMs
two types of transit services~\cite{azure}: one service transits via its private WAN as
much as possible (referred to as the ``\emph{\wan}'' service), and the other uses the public Internet for transit
(referred to as the ``\emph{\inet}'' service). If a VM in Johannesburg uses
the \wan service, Azure will advertise its IP
prefix in both Johannesburg and S\~{a}o Paulo. In contrast, if a VM
uses the \inet service, Azure will
only advertise its IP prefix in Johannesburg. In general, cloud providers expect the private WAN transit to have lower latency.

When we send packets from a RIPE Atlas probe~\cite{RIPE} in Fortaleza, Brazil
to the VM's \wan IP address, we obtain a 254~ms \ac{rtt}.
In contrast, the \ac{rtt} from the same probe to the same VM's \inet IP address is 115~ms.

To understand this latency \textit{inversion}, we examined traceroutes and the network topologies of Azure~\cite{MicrosoftMap} and Angola Cables~\cite{AngolaMap}, as shown by Figure~\ref{fig:network_map}. It turns out that Azure does not own direct submarine cables between South America and Africa, while Angola Cables does.
Thus, when a VM's \inet IP prefix is only advertised in Johannesburg, Angola Cables uses its own submarine cable to deliver the probe's traffic through the South Atlantic Ocean to South Africa,
and traffic enters Azure in Johannesburg, as shown by the solid-line path
in Figure~\ref{fig:case1}.
In contrast, when a VM's \wan IP prefix is advertised in both S\~{a}o Paulo and
Johannesburg, Angola Cables sends the probe's traffic to S\~{a}o
Paulo, resulting in the much longer dashed-line path.


Given that the \wan IP prefix is advertised in both S\~{a}o
Paulo and Johannesburg, why does BGP fail to choose the shorter path?
The route advertised in S\~{a}o Paulo and Johannesburg likely share the same local preferences, since the peering AS at both
locations is Azure. Besides, the AS path
lengths of the two routes are identical.
ISPs often do not honor MEDs~\cite{schlinker2017edge,espresso,Network05-ISPPolice}, so BGP's path selection algorithm relies on interior routing costs to break the tie. In this case, the S\~{a}o Paulo peering location between Azure and Angola Cables is much closer to the probe than the Johannesburg peering
location. Consequently, the traffic takes the early exit at S\~{a}o Paulo, leading to a more circuitous path overall.

%



\begin{figure*}[t]
    \centering
    \includegraphics[width=0.65\linewidth,trim={10pt 10pt 10pt 10pt}]{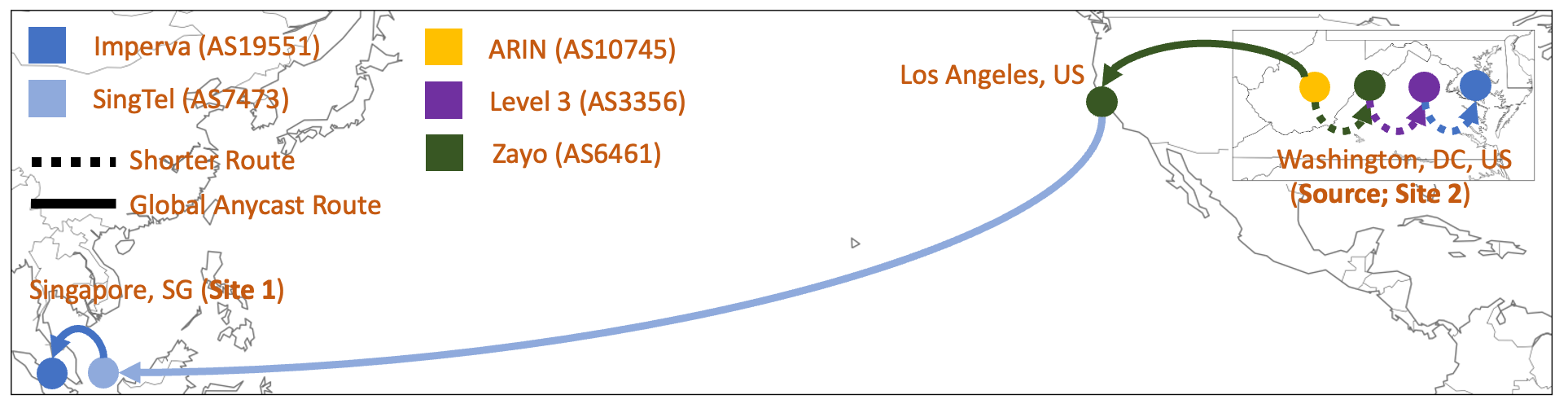}
    \caption{\footnotesize\bfseries A RIPE Atlas probe in Washington DC, US reaches Imperva's site in Singapore via a global anycast IP address, but it can reach the site in Washington DC via a regional anycast IP address. Colors encode path segments in different ASes.\label{fig:case3} }
\end{figure*}

\paragraph{Case 2:} Figure~\ref{fig:case2} shows a Google network detour example.
The \ac{rtt} of the path from a RIPE Atlas probe
in Perth, Australia, towards
a Google VM's \wan IP in Mumbai, India is 196~ms.
However, the \inet path between the same pair has a 149~ms \ac{rtt}.
Using traceroute, we found that the \wan path enters Google's private WAN at Sydney, where the probe's ISP (AARNet) and Google peer, as shown by the dashed line in Figure~\ref{fig:case2}.
On the other hand, for the \inet traffic,  the probe's ISP sends the probe's traffic to Singapore, where it enters the NTT
network and then TATA, which transits the traffic from Singapore to
Mumbai, as shown by the solid line in Figure~\ref{fig:case2}.


In this case, because Google announces the \wan IP prefix in both Mumbai and Sydney, a BGP path to reach the prefix via Singapore is available to the probe's local ISP. However, 
the BGP router at the probe's local ISP fails to
find the shorter path exiting its network in Singapore.
There are two possible reasons: either Google's route is a peer route to AARNet, while NTT's route is a provider route, since NTT is a tier-1 ISP; or AARNet prefers a shorter AS path to reach Google than the longer path that traverses NTT and TATA.



\paragraph{Case 3:} Finally, we show an example observed in the
context of IP anycast~(Figure~\ref{fig:case3}).
In this case, the CDN provider Imperva has sites in Singapore and
Washington DC, US.
It uses regional IP anycast~\cite{SIGCOMM23-RegionalCDN,linkedin-region,Hao2018End-Users} for its CDN service and global IP anycast for its DNS service.
Its regional IP anycast
CDN advertises different regional IP anycast prefixes in Singapore
and Washington, DC, while it
advertises one global IP anycast prefix at both sites for its DNS service.

A RIPE Atlas probe
located in Washington DC reaches Imperva's global anycast address at the Singapore site, resulting in a
250+~ms RTT. In contrast, the same probe reaches the regional anycast
address at the Washington DC site, with a 2~ms RTT.
The reason for this difference is similar to Case 2 above where general BGP policy degrades performance.
The probe's transit provider is Zayo. Imperva's Singapore site uses SingTel as its transit provider, while its Washington DC site uses Level3.
SingTel is Zayo's customer, while Level3 is Zayo's peer.
Because of the prefer-customer routing policy, Zayo prefers the route advertised by SingTel over the route
advertised by Level3 to reach the global IP anycast address.



\section{Latency-Aware BGP}
\label{sec:ASPrep}

\begin{figure}
\includegraphics[width=\linewidth]{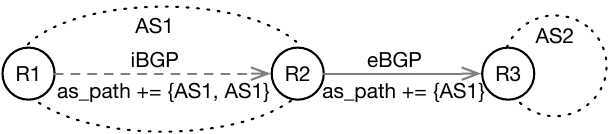}
            \caption{\footnotesize\bfseries An example of prepending AS number to encode the latency information. The arrows refer to messages' direction.\label{fig:as_prepend}}
\end{figure}

%

These examples and others discussed in prior work on cloud routing~\cite{INFOCOM20-PrivateWAN} and anycast~\cite{Li2018Internet, SIGCOMM23-RegionalCDN, rizvi2024anycast} highlight two causes of BGP path inflation: BGP is latency-oblivious, and its policy preferences may lead BGP to select higher latency paths. 
We now explore techniques to address these two shortcomings.

\subsection{Latency-Proportional AS Prepending}
\label{subsec:ASPrep}

\rev{A}{why is ASPATH length signalling the right knob to turn to indicate latency? It seems equally as valid to, say, introduce a new configuration option (what's one more among BGP users), or repurpose another knob like MED}

\rev{B}{applies to all traffic on a prefix rather than just the latency-sensitive traffic. Maybe it's fair to argue that nearly all traffic will benefit from reductions in path latency}

Ideally, BGP's initial design should have included propagation latency (or some equivalent) as a route attribute to permit ASes to make latency-aware route choices. 
Today, such a design change is non-trivial and would require significant modifications to the protocol and to existing implementations. Moreover, exposing path latency could leak information about providers' infrastructure design choices.

\rev{C}{an AS is not incentivized to report their true internal latency, and doing so will degrade the preference of their path if not everyone deploys it.}

\rev{C}{Discussing the potential privacy leak and slight deterioration of BGP-hijack vulnerability. The paper mentions that exposing path latency could leak information about providers’ infrastructure design choices. Wouldn't pretending an AS proportionately to latency do the same?}

We posit that for the relatively few instances of latency inflation that exist in today's Internet (\cref{sec:motivation}), a simpler solution suffices: \textit{encode latency information using AS path prepending}. Specifically, when a path traverses a large AS, the AS prepends its AS number proportional to the latency.


Consider an example in which two ASs, $AS1$ and $AS2$ have an exterior BGP (eBGP) session between their border routers $R2$ and $R3$ (Figure~\ref{fig:as_prepend}). Routers $R1$ and $R2$ in $AS1$
have an interior BGP (iBGP) session. Suppose $R1$ propagates a route advertisement of an IP prefix to $R2$, which in turn exports it to its eBGP neighbor $R3$.

In the current BGP, $R2$ will prepend its AS number ($AS1$) exactly once (unless it chooses to use prepending for traffic engineering) in the AS path attribute of the BGP advertisement when it exports the route to
$R3$. With our proposed modification, $R2$ will prepend its AS number
multiple times in proportion to the \textit{propagation latency} (determined by the underlying infrastructure and speed-of-light constraints) between $R2$ and $R3$. The number of times $R2$ prepends its AS numbers is computed by quantizing the propagation latency:
$\Bigl\lceil \frac{L(R2,R3)}{Q} \Bigr\rceil$,
where $L(R2,R3)$ denotes the latency between $R2$ and $R3$, $Q$ is a quantization factor, and $\lceil x \rceil$ denotes the ceiling integer value of a real number $x$. $R2$ prepends at least once, consistent with the current BGP specification. 

For iBGP, our approach requires \textit{receiver-prepending}: when $R2$ receives the route from $R1$, it prepends its AS
number according to the propagation latency between itself and
$R1$: $\Bigl\lfloor \frac{L(R1,R2)}{Q} \Bigr\rfloor$, where $\lfloor x \rfloor$ denotes the floor integer of $x$.  
Without this, if $R2$ were to use the path length tie-breaker, it would not take latency to $R1$ into account. $R1$ could prepend before advertising to $R2$, but this does not work with route reflection~\cite{RFC4456-RouteReflection}. $R2$ can use the BGP next-hop attribute to determine $R1$'s identity, and prepend accordingly.

Router vendors already support prepending mechanisms, and our approach would require relatively modest implementation changes: the ability to prepend on iBGP sessions, and mechanisms or configurations for latency-proportional prepending. These can be incrementally deployed since they do not require any protocol modifications.



The key challenge in our approach is selecting $Q$ 
to balance latency minimization and the number of different AS path lengths an AS exports to its neighbors. In \S~\ref{sec:simulation}, we use simulations to show how the
quantization factor affects routing overhead and performance. 
Moreover, compared to embedding more granular latency information in a BGP attribute,
quantizing latency exposes fewer details about AS-internal topology and routing, and
it incurs lower advertisement overhead during convergence (\S~\ref{sec:simulation})


\subsection{Local Preference Neutralization}
\label{subsec:localpref}
\rev{B}{The proposal requires reworking business relationships, and it's hard to tell how feasible that would be}

Latency-proportional prepending alone does not suffice, since local preferences can override the AS-path-based selection. For our approach to work, we need to ensure that ASes do not apply local preferences to routes for which they would like to make latency-aware path selection.
We call this \textit{local preference neutralization}.

This is, in general, not possible since local preferences capture business relationships and affect economic choices (\eg, reducing transit costs). We observe that, for the settings we consider (anycast and routing to cloud services) there may exist incentives that permit local preference neutralization. Consider a cloud provider offering a "premium" low-latency tier of service to its cloud customers~\cite{aws,azure,google-traffic-routing}. 
It is likely that, for premium-tier traffic, the cloud provider would be willing to negotiate business agreements with its neighboring ASes to cover both the ``uphill'' and the ``downhill'' transit costs~\cite{gao2021policy} associated with those IP prefixes. In this case,
a provider's neighboring ASes and their customers need not worry about the transit costs for premium-tier IP prefixes. 
They can set the same local preferences to these IP prefixes, so BGP falls back to AS-path-based selection.


Conversely, for outbound traffic from premium-tier IP prefixes, a cloud provider or an IP anycast CDN can neutralize the local preferences to routes learned from all its neighbors, enabling route selection based on the latency-proportional AS path length. This comes at the cost of being unable to minimize transit costs and would apply in cases where revenue from the premium-tier service can offset these costs.

Local preference neutralization does not violate BGP's valley-free
routing policy~\cite{ToN01-GaoRexford,ToN01-ASRel}.
Because valley-free routing is governed by BGP's route export policies (which our approach does not modify), whereas local preferences are set by BGP's import policies (which we propose to modify).

%

Prior work~\cite{SIGCOMM03-PathInflation} has shown that policies contribute relatively little to path inflation; given this, we hypothesize that even when neutralizing local preferences is not feasible, latency-proportional AS prepending can still mitigate path inflation. However, the Internet has become much flatter since that work was conducted, and we will need to validate this hypothesis on the current Internet topology.

\subsection{Putting it Together}

We now revisit the case studies (\cref{sec:example}) to discuss how latency-proportional AS prepending and local preference localization mitigate path inflation in these cases. 
For all three cases, because the latency of the dashed-line
path in Figure~\ref{fig:case1},~\ref{fig:case2}, or~\ref{fig:case3} is
much longer than that of the solid-line path, and each AS would prepend its ASN proportional to its internal latency, the AS path length announced along the dashed-line path will be longer than the one along the solid path.

For Case 1, Angola Cables's BGP router will prefer the solid path even without local preference neutralization, as it has a shorter AS path length, and both the dashed-line and the solid-line routes are advertised directly from Azure.
For Case 2, local preference neutralization eliminates the preference for a peer versus a provider route~(\cref{sec:bgp_select}), so the probe's ISP (AARNET) will equally prefer the dashed-line route advertised by Google and the solid-line route advertised by NTT. Since the latter has a shorter AS path length, it will choose that path. Similarly, for Case 3, after local preference neutralizes Zayo's preference for a customer route from SingTel over a peer route from Level3, Zayo will choose the path to the Washington DC site. This path has the same local preference as the one to the Singapore site but features a shorter AS path length. 


\section{Preliminary Evaluation}\label{sec:simulation}
\rev{C}{Including a discussion on whether the solution can be extended beyond static routing. I believe it is hard to expect BGP to converge before moving away to a less congested path.}

In this section, we present a preliminary simulation-based evaluation on the proposed techniques: latency-proportional AS prepending and local preference neutralization.




We simulate the Internet topology by using CAIDA's router-level Internet topology~\cite{ITDK}. 
This topology provides a router-level view of the Internet, where a node represents a router and an edge denotes the link between two routers detected by traceroutes. 
The topology also specifies the AS that a router belongs to and the geolocations of some routers, but it does not provide the latency of a link between two routers.


We refine the topology to make it applicable to our simulation. First,
as the CAIDA topology does not include latency information, we
estimate an edge's latency using the direct geodesic distance between
two nodes. We use 0.5~ms one-way latency per 100~km distance (namely
1~ms RTT for 100~km distance traveled by light in fiber optics). 
Thus, we only keep the nodes with known geolocations in the topology and
remove the other nodes and involved edges. 
We also remove the nodes without AS specified. 
We retain about 64\% of border routers and the corresponding inter-AS links between them in the topology. 
Second, the CAIDA topology uses border router detection algorithm~\cite{IMC17-bdrmap,IMC18-bdrmapit} and lacks sufficient AS internal link information.
In the simulations,  we assume all border
routers in the same AS are connected to each other. 
In the topology,
we add a BGP route reflector~\cite{RFC4456-RouteReflection} within each AS and  each router is connected to the reflector with
an edge. 
The reflector propagates every iBGP advertisement it receives to all other
routers in the AS. 
We enable the BGP additional
paths option~\cite{RFC7911-BGPAddPath} in the reflector to allow routers in
an AS to select different routes associated with an IP
prefix. Finally, the topology used in the simulations includes
2,930,853 nodes and 6,578,677 edges, covering 58,604 ASes.
 

We implement a customized simulator in Rust that can scale to the Internet topology and simulate three
mechanisms: 1) the original BGP, which we refer to as ``\baseline'';  2) a protocol achieving minimum latency, referred to as ``\optimal'', which simulates the theoretical shortest path algorithm using latency as the link metric; 3) the proposed latency-proportional AS prepending and
local-preference neutralization techniques, referred to as
``\asprep''. 
We experiment with three different values of the \asprep's quantization factor $Q$: 5~ms, 10~ms, and 15~ms.

\begin{figure}[t]
    \centering
    \includegraphics[width=0.83\columnwidth]{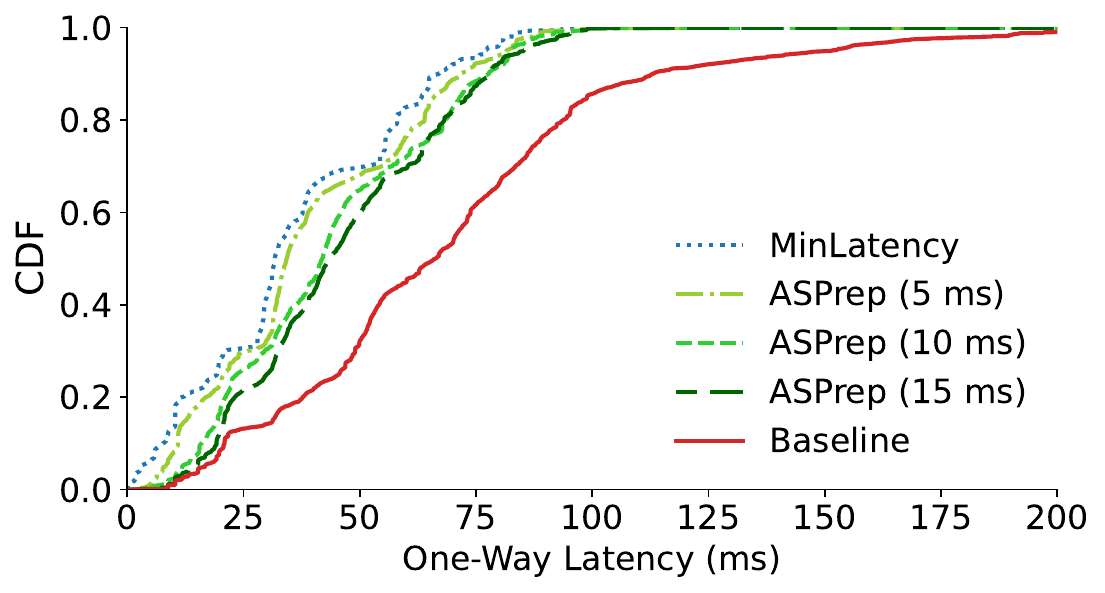}
    \caption{\footnotesize\bfseries CDF of one-way latencies from all nodes to the origin node. The numbers in the legends of \asprep refer to the values of the quantization factor $Q$ we use.\label{fig:cdf_latency}}
\end{figure}

\begin{table}[t]
    \centering
    \begin{tblr}{l|c|c|c}
        \textbf{\quad Protocol} & \textbf{All}        & \textbf{iBGP}       & \textbf{eBGP}    \\
        \hline
        Baseline          & $0.827 \times 10^9$ & $0.822 \times 10^9$ & $5 \times 10^6$  \\
        MinLatency              & $2.302 \times 10^9$ & $2.272 \times 10^9$ & $30 \times 10^6$ \\
        ASPrep (5~ms)     & $2.141 \times 10^9$ & $2.125 \times 10^9$ & $16 \times 10^6$ \\
        ASPrep (10~ms)    & $1.780 \times 10^9$ & $1.768 \times 10^9$ & $12 \times 10^6$ \\
        ASPrep (15~ms)    & $1.262 \times 10^9$ & $1.253 \times 10^9$ & $9 \times 10^6$  \\
    \end{tblr}
    \vspace{8pt}
    \caption{\footnotesize\bfseries The number of routing messages in each
        simulated mechanism. ``All'' refers to the sum of iBGP and eBGP messages.\label{tab:message_cnt}}
\end{table}

In the simulations, we use the AS relationships provided by
CAIDA~\cite{CAIDAASRel,IMC13-ASRel,problink} to implement the export policy
according to \cite{ToN01-ASRel}, and thus we ensure all propagated AS
paths are valley-free. 
We simulate the BGP message propagation when a
BGP router announces a prefix by setting
the origin node as a node of Google (AS~15169) in New York City,
US.
The simulator starts to propagate the BGP updates from this node
to the other nodes. 
When a node receives a route, it will update its
Routing Information Base (RIB) according to the BGP best path
selection algorithm~(\cref{sec:bgp_select}). 
We configure
all ASes to neutralize local preferences for all
simulated cases. 
If the received route replaces the existing one in the RIB, a node will send new BGP updates to its neighbor routers by
iBGP or eBGP. The simulation stops when no new update messages are
generated by any routers, indicating that BGP has
converged~\cite{SIGCOMM99-BGPConvergence,ToN01-GaoRexford}. 
After the
convergence, the simulator calculates the one-way latency from each
node to the origin node in the topology by summing up the latency of
each edge in the path. 
The simulator also counts the generated update
messages during the convergence process.



Figure~\ref{fig:cdf_latency} shows the CDF of the latency to the
origin node from all nodes in the topology.
As we can see, there is a wide gap between \baseline and the other cases, indicating that both \asprep and \optimal achieve significant latency reduction. For instance, \asprep~(15~ms) reduces the 90th percentile latency from \baseline's 113~ms to 78~ms (31\% off). Besides, as $Q$ increases, the overall latency of ASPrep
increases slightly, but it still improves the latency significantly compared to
\baseline even when $Q=15$~ms.  The median latencies of the simulated
protocols are 31.7~ms (\optimal), 34.3~ms (\asprep, 5~ms), 41.9~ms
(\asprep, 10~ms), 43.4 (\asprep, 15~ms), and 66.3~ms (\baseline),
respectively. Overall, \asprep's performance lies in-between \optimal and
\baseline and is closer to \optimal for smaller $Q$ values.


\Cref{tab:message_cnt} presents the message counts each protocol uses to converge.  
We note that the simulation's message count is a theoretical upper bound as we have not implemented
route flap damping~\cite{RFC2439-RouteDamping} or Minimum Route
Advertisement Interval (MRAI)~\cite{RFC4098-Term} and all ASes use route reflectors.  Overall, the
\baseline case has the fewest number of messages since all possible
internal paths within an AS to the simulated prefix are represented
with one AS number prepending. The \optimal case takes about $2.8
\times$ more messages (2.3~billion vs 0.8~billion) than the \baseline
case, because the BGP routes it propagates include the latency
information of every AS internal path. This explosion of internal
paths enables the \optimal protocol to select the minimal latency path
at the cost of routing overhead.  The message count of \asprep is
close to \optimal when $Q=5$~ms, but it reduces significantly when $Q$
increases.  For $Q=15$~ms, \asprep takes only 55\% of \optimal's message count to converge and about $1.5 \times$ messages of
\baseline. Similar to \optimal, it exposes the latency information in the
route advertisement, but it quantizes the latency to reduce the message count. 
Therefore,  when
the latency difference between the newly received route and the
existing route is small, the node will not update its RIB or
disseminate new update messages. 
The iBGP message counts are a
few magnitudes larger than eBGP message counts, because, when a node
receives an iBGP message and updates its RIB, it needs to propagate
the update to all other nodes in the same AS through the route
reflector.



In conclusion, we observe that latency-proportional AS prepending and
local preference neutralization achieve a good tradeoff between
latency minimization and routing overhead.
Although preliminary, the simulation-based evaluation shows promising results for the proposed techniques, warranting further investigation.

 \section{Future Directions}

The open topics we plan to investigate in the future are:

\rev{A}{Other than a brief one-paragraph discussion in S5, the paper doesn't discuss partial deployment, which seems especially important for a proposal like this, }

\paragraph{Partial Deployment:} Our simulation study assumes
all ASes deploy the proposed BGP modifications. The impacts of
a partial deployment, where a subset of ASes and their neighbors
deploy latency-proportional AS prepending and/or local preference
neutralization remains unclear.  Additionally, it is not clear what 
fraction of ASes is needed to bootstrap the deployment for achieving
sufficient benefits for further
deployments~\cite{adoptability-model,raja2017dbgp}. 
We plan to conduct more extensive simulations and analysis to answer these questions. 

\paragraph{Quantization Factor and Overhead:} The simulation
results demonstrate that a coarser quantization factor leads to lower
routing overhead. 
The precise scaling relationship between these two should be investigated to optimize the performance and overhead tradeoff.

\paragraph{Routing Dynamics:} The current simulation study uses
a static Internet topology and does not consider routing
dynamics. Questions such as 
how latency-proportional AS
prepending will affect BGP convergence time and routing message
overhead require future experiments and analysis.

\paragraph{Real-world Validation and Implementation} Lastly, although we obtain a state-of-the-art Internet
topology, it is still inaccurate and
does not fully capture the complexity of the Internet.
Additionally, the specific router configurations or software changes
required for deploying the proposed techniques are unclear.  We plan
to implement the proposed changes on an open-source router
implementation such as BIRD~\cite{bird} and conduct real-world
experiments either using a research testbed (PEERING~\cite{schlinker19peering} or
Tangled~\cite{bertholdo-tangled-IM21}) or a cloud provider such as Vultr~\cite{vultr} that
allows a tenant to announce its own address. These will deepen our understanding of
the costs and benefits of the proposed techniques.


\section{Related Work}

%
\paragraph{Geo-hint:} After observing the path inflation problem in the global IP anycast system, Li et al.~\cite{Li2018Internet} proposed to add geo-hints to BGP route advertisements. 
A BGP router can use the geo-hints to choose a geographically close anycast replica. 
This technique can effectively route anycast traffic to the geographically closest replica, but cannot address Cases 1 and 2 path inflation problems. 
In these two examples, there is only one VM location. A BGP router cannot use geodesic distance to the VM to differentiate the two routes it receives. Additionally, it requires router vendors to change BGP implementation to account for the geo-hint attribute.


\paragraph{Accumulated Interior Gateway Protocol (AIGP) Metric:} RFC 7311~\cite{RFC7311-AIGP} introduced the AIGP metric to enable BGP networks under a single administration to use this metric for path selection.  As its name implies, this metric accumulates the IGP costs across multiple ASes. 
A BGP speaker in the ASes under the same administration compares this attribute after the local preference attribute and before all other tie-breaking attributes such as AS path lengths.
If all ASes set their IGP costs according to link latency and
accumulate their IGP costs in BGP messages, then AIGP under this
setting becomes the \optimal protocol we simulate (\cref{sec:simulation}) without local preference
neutralization. Although it can avoid path inflation, it exposes 
detailed AS internal route metrics to the broader Internet and introduces
high routing overhead. AIGP is a current standard enabled
by router vendors, but its RFC explicitly specifies that it should not
be enabled between networks under different administrations.

%
%

\paragraph{BGP Engineering:} PAINTER~\cite{SIGCOMM23-PAINTER} and Tango~\cite{tango-hotnets,tango-usenix} expose multiple AS-level routes to edge networks by advertising different IP prefixes to different network providers. Edge networks can select a BGP path with minimal latency by testing different addresses.  
DailyCatch~\cite{IMC19-Taming} uses periodic measurements to select the set of neighboring ASes to announce an IP anycast prefix in order to achieve low anycast latency.
These solutions, if deployed, can mitigate BGP path inflation, but
deploying them requires extensive trial-and-error engineering
work. PAINTER and Tango also require additional IP address
space and edge network deployments. The techniques proposed in our work address the
fundamental causes of BGP path inflation.


\section{Conclusion}
\label{sec:conclusion}

We propose two simple modifications to BGP to address its
long-standing path inflation problem. The first modification is 
latency-proportional AS path prepending: an AS prepends itself in a BGP route proportionally to 
its transit path length. The second
modification is to enable ASes to set the same local preferences for
high-value latency-sensitive routes to enable latency-based path
selection. 
A preliminary simulation study on an Internet-scale
topology indicates that these two techniques are promising and can
achieve a good tradeoff between latency minimization and routing
overhead.

%



\bibliographystyle{ACM-Reference-Format}
\bibliography{reference}


\begin{thebibliography}{59}


\ifx \showCODEN    \undefined \def \showCODEN     #1{\unskip}     \fi
\ifx \showDOI      \undefined \def \showDOI       #1{#1}\fi
\ifx \showISBNx    \undefined \def \showISBNx     #1{\unskip}     \fi
\ifx \showISBNxiii \undefined \def \showISBNxiii  #1{\unskip}     \fi
\ifx \showISSN     \undefined \def \showISSN      #1{\unskip}     \fi
\ifx \showLCCN     \undefined \def \showLCCN      #1{\unskip}     \fi
\ifx \shownote     \undefined \def \shownote      #1{#1}          \fi
\ifx \showarticletitle \undefined \def \showarticletitle #1{#1}   \fi
\ifx \showURL      \undefined \def \showURL       {\relax}        \fi
\providecommand\bibfield[2]{#2}
\providecommand\bibinfo[2]{#2}
\providecommand\natexlab[1]{#1}
\providecommand\showeprint[2][]{arXiv:#2}

\bibitem[Ang({[n.\,d.]})]%
        {AngolaMap}
 \bibinfo{year}{[n.\,d.]}\natexlab{}.
\newblock \bibinfo{title}{{Angola Cables Network Map}}.
\newblock \bibinfo{howpublished}{Retrieved in June, 2024 from \url{https://angolacables.co.ao/en/rede-global/}}.
\newblock


\bibitem[CAI({[n.\,d.]})]%
        {CAIDAASRel}
 \bibinfo{year}{[n.\,d.]}\natexlab{}.
\newblock \bibinfo{title}{{AS Relationships}}.
\newblock \bibinfo{howpublished}{Retrieved in Dec, 2022 from \url{https://www.caida.org/catalog/datasets/as-relationships/}}.
\newblock


\bibitem[aws({[n.\,d.]})]%
        {aws}
 \bibinfo{year}{[n.\,d.]}\natexlab{}.
\newblock \bibinfo{title}{{AWS Global Accelerator Features}}.
\newblock \bibinfo{howpublished}{Retrieved in April, 2024 from \url{https://aws.amazon.com/global-accelerator/features/}}.
\newblock


\bibitem[Mic({[n.\,d.]})]%
        {MicrosoftMap}
 \bibinfo{year}{[n.\,d.]}\natexlab{}.
\newblock \bibinfo{title}{{Azure Global Infrastructure Experience}}.
\newblock \bibinfo{howpublished}{Retrieved in June, 2024 from \url{https://datacenters.microsoft.com/globe/explore/}}.
\newblock


\bibitem[ITD({[n.\,d.]})]%
        {ITDK}
 \bibinfo{year}{[n.\,d.]}\natexlab{}.
\newblock \bibinfo{title}{{Macroscopic Internet Topology Data Kit (ITDK)}}.
\newblock \bibinfo{howpublished}{Retrieved in Dec, 2022 from \url{https://www.caida.org/catalog/datasets/internet-topology-data-kit/}}.
\newblock


\bibitem[bir({[n.\,d.]})]%
        {bird}
 \bibinfo{year}{[n.\,d.]}\natexlab{}.
\newblock \bibinfo{title}{{The BIRD Internet Routing Daemon}}.
\newblock
\newblock
\newblock
\shownote{\url{http://bird.network.cz/}}.


\bibitem[vul({[n.\,d.]})]%
        {vultr}
 \bibinfo{year}{[n.\,d.]}\natexlab{}.
\newblock \bibinfo{title}{{Vultr: Announce Your Own IP Space}}.
\newblock
\newblock
\newblock
\shownote{\url{https://www.vultr.com/features/bgp/}}.


\bibitem[azu({[n.\,d.]})]%
        {azure}
 \bibinfo{year}{[n.\,d.]}\natexlab{}.
\newblock \bibinfo{title}{{What is Routing Preference?}}
\newblock \bibinfo{howpublished}{Retrieved in April, 2024 from \url{https://learn.microsoft.com/en-us/azure/virtual-network/ip-services/routing-preference-overview}}.
\newblock


\bibitem[Andersen et~al\mbox{.}(2001)]%
        {resilient-andersen-2001}
\bibfield{author}{\bibinfo{person}{David Andersen}, \bibinfo{person}{Hari Balakrishnan}, \bibinfo{person}{Frans Kaashoek}, {and} \bibinfo{person}{Robert Morris}.} \bibinfo{year}{2001}\natexlab{}.
\newblock \showarticletitle{{Resilient Overlay Networks}}. In \bibinfo{booktitle}{\emph{Proceedings of the eighteenth ACM symposium on Operating systems principles (SOSP'01)}}.
\newblock


\bibitem[Arnold et~al\mbox{.}(2019)]%
        {beatbgp-todd-hotnet18}
\bibfield{author}{\bibinfo{person}{Todd Arnold}, \bibinfo{person}{Matt Calder}, \bibinfo{person}{Italo Cunha}, \bibinfo{person}{Arpit Gupta}, \bibinfo{person}{Harsha~V. Madhyastha}, \bibinfo{person}{Michael Schapira}, {and} \bibinfo{person}{Ethan Katz-Bassett}.} \bibinfo{year}{2019}\natexlab{}.
\newblock \showarticletitle{Beating BGP is Harder than we Thought}. In \bibinfo{booktitle}{\emph{Proceedings of HotNets}}.
\newblock


\bibitem[Arnold et~al\mbox{.}(2020)]%
        {INFOCOM20-PrivateWAN}
\bibfield{author}{\bibinfo{person}{Todd Arnold}, \bibinfo{person}{Ege G{\"u}rmeri{\c c}liler}, \bibinfo{person}{Georgia Essig}, \bibinfo{person}{Arpit Gupta}, \bibinfo{person}{Matt Calder}, \bibinfo{person}{Vasileios Giotsas}, {and} \bibinfo{person}{Ethan Katz-Bassett}.} \bibinfo{year}{2020}\natexlab{}.
\newblock \showarticletitle{{(How Much) Does a Private WAN Improve Cloud Performance?}}. In \bibinfo{booktitle}{\emph{Proceedings of INFOCOM}}. IEEE.
\newblock


\bibitem[Azure(2024)]%
        {azure-pops}
\bibfield{author}{\bibinfo{person}{Azure}.} \bibinfo{year}{2024}\natexlab{}.
\newblock \bibinfo{title}{Azure Front Door POP locations by metro}.
\newblock \bibinfo{howpublished}{Retrieved in Jan, 2024 from \url{https://learn.microsoft.com/en-us/azure/frontdoor/edge-locations-by-region}}.
\newblock


\bibitem[Bates et~al\mbox{.}(2006)]%
        {RFC4456-RouteReflection}
\bibfield{author}{\bibinfo{person}{Tony Bates}, \bibinfo{person}{Enke Chen}, {and} \bibinfo{person}{Ravi Chandra}.} \bibinfo{year}{2006}\natexlab{}.
\newblock \bibinfo{booktitle}{\emph{{RFC 4456: BGP Route Reflection: an Alternative to Full Mesh Internal BGP (IBGP)}}}.
\newblock \bibinfo{type}{Internet Requests for Comments}. \bibinfo{institution}{Internet Engineering Task Force (IETF)}.
\newblock


\bibitem[Berkowitz et~al\mbox{.}(2005)]%
        {RFC4098-Term}
\bibfield{author}{\bibinfo{person}{H Berkowitz}, \bibinfo{person}{S Hares}, \bibinfo{person}{P Krishnaswamy}, {and} \bibinfo{person}{M Lepp}.} \bibinfo{year}{2005}\natexlab{}.
\newblock \bibinfo{booktitle}{\emph{{RFC 4098: Terminology for Benchmarking BGP Device Convergence in the Control Plane}}}.
\newblock \bibinfo{type}{Internet Requests for Comments}. \bibinfo{institution}{Internet Engineering Task Force (IETF)}.
\newblock


\bibitem[Bertholdo et~al\mbox{.}(2021)]%
        {bertholdo-tangled-IM21}
\bibfield{author}{\bibinfo{person}{Leandro~M Bertholdo}, \bibinfo{person}{Joao~M Ceron}, \bibinfo{person}{Wouter~B de Vries}, \bibinfo{person}{Ricardo de Oliveira~Schmidt}, \bibinfo{person}{Lisandro~Zambenedetti Granville}, \bibinfo{person}{Roland van Rijswijk-Deij}, {and} \bibinfo{person}{Aiko Pras}.} \bibinfo{year}{2021}\natexlab{}.
\newblock \showarticletitle{{Tangled: A Cooperative Anycast Testbed}}. In \bibinfo{booktitle}{\emph{Proceedings of the IFIP/IEEE International Symposium on Integrated Network Management}}. IEEE.
\newblock


\bibitem[Bhattacherjee et~al\mbox{.}(2022)]%
        {NSDI22-cisp}
\bibfield{author}{\bibinfo{person}{Debopam Bhattacherjee}, \bibinfo{person}{Waqar Aqeel}, \bibinfo{person}{Sangeetha~Abdu Jyothi}, \bibinfo{person}{Ilker~Nadi Bozkurt}, \bibinfo{person}{William Sentosa}, \bibinfo{person}{Muhammad Tirmazi}, \bibinfo{person}{Anthony Aguirre}, \bibinfo{person}{Balakrishnan Chandrasekaran}, \bibinfo{person}{P~Brighten Godfrey}, \bibinfo{person}{Gregory Laughlin}, {et~al\mbox{.}}} \bibinfo{year}{2022}\natexlab{}.
\newblock \showarticletitle{{cISP: A Speed-of-Light Internet Service Provider}}. In \bibinfo{booktitle}{\emph{Proceedings of NSDI}}. USENIX, \bibinfo{pages}{1115--1133}.
\newblock


\bibitem[Birge-Lee et~al\mbox{.}(2022)]%
        {tango-hotnets}
\bibfield{author}{\bibinfo{person}{Henry Birge-Lee}, \bibinfo{person}{Maria Apostolaki}, {and} \bibinfo{person}{Jennifer Rexford}.} \bibinfo{year}{2022}\natexlab{}.
\newblock \showarticletitle{It takes two to tango: cooperative edge-to-edge routing}. In \bibinfo{booktitle}{\emph{Proceedings of HotNets}}. ACM.
\newblock


\bibitem[Birge-Lee et~al\mbox{.}(2024)]%
        {tango-usenix}
\bibfield{author}{\bibinfo{person}{Henry Birge-Lee}, \bibinfo{person}{Sophia Yoo}, \bibinfo{person}{Benjamin Herber}, \bibinfo{person}{Jennifer Rexford}, {and} \bibinfo{person}{Maria Apostolaki}.} \bibinfo{year}{2024}\natexlab{}.
\newblock \showarticletitle{{TANGO: Secure Collaborative Route Control across the Public Internet}}. In \bibinfo{booktitle}{\emph{Proceedings of NSDI 24}}. USENIX.
\newblock


\bibitem[Caesar and Rexford(2005)]%
        {Network05-ISPPolice}
\bibfield{author}{\bibinfo{person}{Matthew Caesar} {and} \bibinfo{person}{Jennifer Rexford}.} \bibinfo{year}{2005}\natexlab{}.
\newblock \showarticletitle{{BGP Routing Policies in ISP Networks}}.
\newblock \bibinfo{journal}{\emph{IEEE network}} \bibinfo{volume}{19}, \bibinfo{number}{6} (\bibinfo{year}{2005}), \bibinfo{pages}{5--11}.
\newblock


\bibitem[Chan et~al\mbox{.}(2006)]%
        {adoptability-model}
\bibfield{author}{\bibinfo{person}{Haowen Chan}, \bibinfo{person}{Debabrata Dash}, \bibinfo{person}{Adrian Perrig}, {and} \bibinfo{person}{Hui Zhang}.} \bibinfo{year}{2006}\natexlab{}.
\newblock \showarticletitle{{Modeling adoptability of secure BGP protocol}}. In \bibinfo{booktitle}{\emph{Proceedings of SIGCOMM}}. ACM.
\newblock


\bibitem[Chen et~al\mbox{.}(2015)]%
        {Chen2015End}
\bibfield{author}{\bibinfo{person}{Fangfei Chen}, \bibinfo{person}{Ramesh~K. Sitaraman}, {and} \bibinfo{person}{Marcelo Torres}.} \bibinfo{year}{2015}\natexlab{}.
\newblock \showarticletitle{{End-User Mapping: Next Generation Request Routing for Content Delivery}}. In \bibinfo{booktitle}{\emph{Proceedings of SIGCOMM}}. ACM.
\newblock


\bibitem[Chuat et~al\mbox{.}(2022)]%
        {chuat2022complete}
\bibfield{author}{\bibinfo{person}{Laurent Chuat}, \bibinfo{person}{Markus Legner}, \bibinfo{person}{David Basin}, \bibinfo{person}{David Hausheer}, \bibinfo{person}{Samuel Hitz}, \bibinfo{person}{Peter M{\"u}ller}, {and} \bibinfo{person}{Adrian Perrig}.} \bibinfo{year}{2022}\natexlab{}.
\newblock \showarticletitle{{The Complete Guide to SCION}}.
\newblock \bibinfo{journal}{\emph{Information Security and Cryptography}} (\bibinfo{year}{2022}).
\newblock


\bibitem[Cloud({[n.\,d.]})]%
        {google-traffic-routing}
\bibfield{author}{\bibinfo{person}{Google Cloud}.} \bibinfo{year}{[n.\,d.]}\natexlab{}.
\newblock \bibinfo{title}{Network Service Tiers overview}.
\newblock \bibinfo{howpublished}{Retrieved in May, 2024 from \url{https://cloud.google.com/network-tiers/docs/overview\#traffic_routing}}.
\newblock


\bibitem[De~Vries et~al\mbox{.}(2019)]%
        {de-GDNS-19TNS}
\bibfield{author}{\bibinfo{person}{Wouter~B De~Vries}, \bibinfo{person}{Roland van Rijswijk-Deij}, \bibinfo{person}{Pieter-Tjerk De~Boer}, {and} \bibinfo{person}{Aiko Pras}.} \bibinfo{year}{2019}\natexlab{}.
\newblock \showarticletitle{{Passive Observations of a Large DNS Service: 2.5 Years in the Life of Google}}.
\newblock \bibinfo{journal}{\emph{IEEE Transactions on Network and Service Management}} \bibinfo{volume}{17}, \bibinfo{number}{1} (\bibinfo{year}{2019}), \bibinfo{pages}{190--200}.
\newblock


\bibitem[Gao(2001)]%
        {ToN01-ASRel}
\bibfield{author}{\bibinfo{person}{Lixin Gao}.} \bibinfo{year}{2001}\natexlab{}.
\newblock \showarticletitle{{On Inferring Autonomous System Relationships in the Internet}}.
\newblock \bibinfo{journal}{\emph{IEEE/ACM Transactions on Networking}} \bibinfo{volume}{9}, \bibinfo{number}{6} (\bibinfo{year}{2001}), \bibinfo{pages}{733--745}.
\newblock


\bibitem[Gao and Rexford(2001)]%
        {ToN01-GaoRexford}
\bibfield{author}{\bibinfo{person}{Lixin Gao} {and} \bibinfo{person}{Jennifer Rexford}.} \bibinfo{year}{2001}\natexlab{}.
\newblock \showarticletitle{{Stable Internet Routing without Global Coordination}}.
\newblock \bibinfo{journal}{\emph{IEEE/ACM Transactions on Networking}} \bibinfo{volume}{9}, \bibinfo{number}{6} (\bibinfo{year}{2001}), \bibinfo{pages}{681--692}.
\newblock


\bibitem[Griffin and Wilfong(1999)]%
        {SIGCOMM99-BGPConvergence}
\bibfield{author}{\bibinfo{person}{Timothy~G Griffin} {and} \bibinfo{person}{Gordon Wilfong}.} \bibinfo{year}{1999}\natexlab{}.
\newblock \showarticletitle{{An Analysis of BGP Convergence Properties}}. In \bibinfo{booktitle}{\emph{Proceedings of SIGCOMM}}. ACM, \bibinfo{pages}{277--288}.
\newblock


\bibitem[Hao et~al\mbox{.}(2018)]%
        {Hao2018End-Users}
\bibfield{author}{\bibinfo{person}{Shuai Hao}, \bibinfo{person}{Yubao Zhang}, \bibinfo{person}{Haining Wang}, {and} \bibinfo{person}{Angelos Stavrou}.} \bibinfo{year}{2018}\natexlab{}.
\newblock \showarticletitle{{End-Users Get Maneuvered: Empirical Analysis of Redirection Hijacking in Content Delivery Networks}}. In \bibinfo{booktitle}{\emph{Proceedings of 27th USENIX Security Symposium (USENIX Security'18)}}.
\newblock


\bibitem[Jin et~al\mbox{.}(2019)]%
        {problink}
\bibfield{author}{\bibinfo{person}{Yuchen Jin}, \bibinfo{person}{Colin Scott}, \bibinfo{person}{Amogh Dhamdhere}, \bibinfo{person}{Vasileios Giotsas}, \bibinfo{person}{Arvind Krishnamurthy}, {and} \bibinfo{person}{Scott Shenker}.} \bibinfo{year}{2019}\natexlab{}.
\newblock \showarticletitle{Stable and Practical {AS} Relationship Inference with {ProbLink}}. In \bibinfo{booktitle}{\emph{Proceedings of NSDI}}. USENIX.
\newblock


\bibitem[Koch et~al\mbox{.}(2021)]%
        {twotale-li-sigcomm21}
\bibfield{author}{\bibinfo{person}{Thomas Koch}, \bibinfo{person}{Ke Li}, \bibinfo{person}{Calvin Ardi}, \bibinfo{person}{Ethan Katz-Bassett}, \bibinfo{person}{Matt Calder}, {and} \bibinfo{person}{John Heidemann}.} \bibinfo{year}{2021}\natexlab{}.
\newblock \showarticletitle{{Anycast in Context: A Tale of Two Systems}}. In \bibinfo{booktitle}{\emph{Proceedings of the Annual Conference of the Conference of the ACM Special Interest Group on Data Communication (SIGCOMM'21)}}. ACM.
\newblock


\bibitem[Koch et~al\mbox{.}(2023)]%
        {SIGCOMM23-PAINTER}
\bibfield{author}{\bibinfo{person}{Thomas Koch}, \bibinfo{person}{Shuyue Yu}, \bibinfo{person}{Sharad Agarwal}, \bibinfo{person}{Ethan Katz-Bassett}, {and} \bibinfo{person}{Ryan Beckett}.} \bibinfo{year}{2023}\natexlab{}.
\newblock \showarticletitle{PAINTER: Ingress Traffic Engineering and Routing for Enterprise Cloud Networks}. In \bibinfo{booktitle}{\emph{Proceedings of SIGCOMM}}. ACM.
\newblock


\bibitem[Kuipers({[n.\,d.]})]%
        {JH2015Analysing}
\bibfield{author}{\bibinfo{person}{JH Kuipers}.} \bibinfo{year}{[n.\,d.]}\natexlab{}.
\newblock \bibinfo{title}{{Analysing the K-root Anycast Infrastructure}}.
\newblock \bibinfo{howpublished}{Retrieved in Sep, 2022 from \url{https://labs.ripe.net/Members/jh_kuipers/analyzing-the-k-root-anycast-infrastructure}}.
\newblock


\bibitem[Li et~al\mbox{.}(2018)]%
        {Li2018Internet}
\bibfield{author}{\bibinfo{person}{Zhihao Li}, \bibinfo{person}{Dave Levin}, \bibinfo{person}{Neil Spring}, {and} \bibinfo{person}{Bobby Bhattacharjee}.} \bibinfo{year}{2018}\natexlab{}.
\newblock \showarticletitle{{Internet Anycast: Performance, Problems, \& Potential}}. In \bibinfo{booktitle}{\emph{Proceedings of SIGCOMM}}. ACM.
\newblock


\bibitem[Liu et~al\mbox{.}(2007)]%
        {iLiu2007Two}
\bibfield{author}{\bibinfo{person}{Ziqian Liu}, \bibinfo{person}{Bradley Huffaker}, \bibinfo{person}{Marina Fomenkov}, \bibinfo{person}{Nevil Brownlee}, {and} \bibinfo{person}{Kimberly Claffy}.} \bibinfo{year}{2007}\natexlab{}.
\newblock \showarticletitle{{Two Days in the Life of the DNS Anycast Root Servers}}. In \bibinfo{booktitle}{\emph{Proceedings of PAM}}. Springer.
\newblock


\bibitem[Luckie et~al\mbox{.}(2016)]%
        {IMC17-bdrmap}
\bibfield{author}{\bibinfo{person}{Matthew Luckie}, \bibinfo{person}{Amogh Dhamdhere}, \bibinfo{person}{Bradley Huffaker}, \bibinfo{person}{David Clark}, {and} \bibinfo{person}{KC Claffy}.} \bibinfo{year}{2016}\natexlab{}.
\newblock \showarticletitle{{bdrmap: Inference of Borders between IP Networks}}. In \bibinfo{booktitle}{\emph{Proceedings of IMC}}. ACM, \bibinfo{pages}{381--396}.
\newblock


\bibitem[Luckie et~al\mbox{.}(2013)]%
        {IMC13-ASRel}
\bibfield{author}{\bibinfo{person}{Matthew Luckie}, \bibinfo{person}{Bradley Huffaker}, \bibinfo{person}{Amogh Dhamdhere}, \bibinfo{person}{Vasileios Giotsas}, {and} \bibinfo{person}{KC Claffy}.} \bibinfo{year}{2013}\natexlab{}.
\newblock \showarticletitle{{AS Relationships, Customer cones, and Validation}}. In \bibinfo{booktitle}{\emph{Proceedings of IMC}}. ACM, \bibinfo{pages}{243--256}.
\newblock


\bibitem[Maheshwari(2015)]%
        {linkedin-region}
\bibfield{author}{\bibinfo{person}{Ritesh Maheshwari}.} \bibinfo{year}{2015}\natexlab{}.
\newblock \bibinfo{title}{{TCP over IP Anycast - Pipe Dream or Reality?}}
\newblock \bibinfo{howpublished}{Retrieved in Sep, 2022 from \url{https://engineering.linkedin.com/network-performance/tcp-over-ip-anycast-pipe-dream-or-reality}}.
\newblock


\bibitem[Marder et~al\mbox{.}(2018)]%
        {IMC18-bdrmapit}
\bibfield{author}{\bibinfo{person}{Alexander Marder}, \bibinfo{person}{Matthew Luckie}, \bibinfo{person}{Amogh Dhamdhere}, \bibinfo{person}{Bradley Huffaker}, \bibinfo{person}{KC Claffy}, {and} \bibinfo{person}{Jonathan~M Smith}.} \bibinfo{year}{2018}\natexlab{}.
\newblock \showarticletitle{Pushing the boundaries with bdrmapit: Mapping router ownership at internet scale}. In \bibinfo{booktitle}{\emph{Proceedings of IMC}}. ACM, \bibinfo{pages}{56--69}.
\newblock


\bibitem[McCauley et~al\mbox{.}(2019)]%
        {trotsky-sigcomm19}
\bibfield{author}{\bibinfo{person}{James McCauley}, \bibinfo{person}{Yotam Harchol}, \bibinfo{person}{Aurojit Panda}, \bibinfo{person}{Barath Raghavan}, {and} \bibinfo{person}{Scott Shenker}.} \bibinfo{year}{2019}\natexlab{}.
\newblock \showarticletitle{{Enabling a Permanent Revolution in Internet Architecture}}. In \bibinfo{booktitle}{\emph{SIGCOMM'19}}.
\newblock


\bibitem[McQuistin et~al\mbox{.}(2019)]%
        {IMC19-Taming}
\bibfield{author}{\bibinfo{person}{Stephen McQuistin}, \bibinfo{person}{Sree~Priyanka Uppu}, {and} \bibinfo{person}{Marcel Flores}.} \bibinfo{year}{2019}\natexlab{}.
\newblock \showarticletitle{{Taming Anycast in the Wild Internet}}. In \bibinfo{booktitle}{\emph{Proceedings of IMC}}. ACM.
\newblock


\bibitem[Mohan et~al\mbox{.}(2020)]%
        {hotnets-edge-computing}
\bibfield{author}{\bibinfo{person}{Nitinder Mohan}, \bibinfo{person}{Lorenzo Corneo}, \bibinfo{person}{Aleksandr Zavodovski}, \bibinfo{person}{Suzan Bayhan}, \bibinfo{person}{Walter Wong}, {and} \bibinfo{person}{Jussi Kangasharju}.} \bibinfo{year}{2020}\natexlab{}.
\newblock \showarticletitle{{Pruning Edge Research with Latency Shears}}. In \bibinfo{booktitle}{\emph{Proceedings of HotNets}}. ACM.
\newblock


\bibitem[Mohapatra et~al\mbox{.}(2014)]%
        {RFC7311-AIGP}
\bibfield{author}{\bibinfo{person}{Prodosh Mohapatra}, \bibinfo{person}{Rex Fernando}, \bibinfo{person}{Eric~C. Rosen}, {and} \bibinfo{person}{Jim Uttaro}.} \bibinfo{year}{2014}\natexlab{}.
\newblock \bibinfo{booktitle}{\emph{{RFC 7311: The Accumulated IGP Metric Attribute for BGP}}}.
\newblock \bibinfo{type}{Internet Requests for Comments}. \bibinfo{institution}{Internet Engineering Task Force (IETF)}.
\newblock


\bibitem[Moy(1998)]%
        {RFC2328-OSPF}
\bibfield{author}{\bibinfo{person}{John Moy}.} \bibinfo{year}{1998}\natexlab{}.
\newblock \bibinfo{booktitle}{\emph{RFC 2328: OSPF Version 2}}.
\newblock \bibinfo{type}{Internet Requests for Comments}. \bibinfo{institution}{Internet Engineering Task Force (IETF)}.
\newblock


\bibitem[Oran(1990)]%
        {RFC1142-ISIS}
\bibfield{author}{\bibinfo{person}{David Oran}.} \bibinfo{year}{1990}\natexlab{}.
\newblock \bibinfo{booktitle}{\emph{RFC 1142: OSI IS-IS intra-domain routing protocol}}.
\newblock \bibinfo{type}{Internet Requests for Comments}. \bibinfo{institution}{Internet Engineering Task Force (IETF)}.
\newblock


\bibitem[Peter et~al\mbox{.}(2014)]%
        {Arrow}
\bibfield{author}{\bibinfo{person}{Simon Peter}, \bibinfo{person}{Umar Javed}, \bibinfo{person}{Qiao Zhang}, \bibinfo{person}{Doug Woos}, \bibinfo{person}{Thomas Anderson}, {and} \bibinfo{person}{Arvind Krishnamurthy}.} \bibinfo{year}{2014}\natexlab{}.
\newblock \showarticletitle{{One Tunnel is (Often) Enough}}. In \bibinfo{booktitle}{\emph{Proceedings of SIGCOMM}}. ACM, \bibinfo{pages}{99–110}.
\newblock


\bibitem[Rekhter and Li(1994)]%
        {rfc4271_bgp}
\bibfield{author}{\bibinfo{person}{Yakov Rekhter} {and} \bibinfo{person}{Tony Li}.} \bibinfo{year}{1994}\natexlab{}.
\newblock \bibinfo{booktitle}{\emph{{RFC 4271: A Border Gateway Protocol 4 (BGP-4)}}}.
\newblock \bibinfo{type}{Internet Requests for Comments}. \bibinfo{institution}{Internet Engineering Task Force (IETF)}.
\newblock


\bibitem[Rizvi et~al\mbox{.}(2024)]%
        {rizvi2024anycast}
\bibfield{author}{\bibinfo{person}{ASM Rizvi}, \bibinfo{person}{Tingshan Huang}, \bibinfo{person}{Rasit Esrefoglu}, {and} \bibinfo{person}{John Heidemann}.} \bibinfo{year}{2024}\natexlab{}.
\newblock \showarticletitle{Anycast Polarization in the Wild}. In \bibinfo{booktitle}{\emph{Proceedings of PAM}}.
\newblock


\bibitem[Sambasivan et~al\mbox{.}(2017)]%
        {raja2017dbgp}
\bibfield{author}{\bibinfo{person}{Raja~R. Sambasivan}, \bibinfo{person}{David Tran-Lam}, \bibinfo{person}{Aditya Akella}, {and} \bibinfo{person}{Peter Steenkiste}.} \bibinfo{year}{2017}\natexlab{}.
\newblock \showarticletitle{{Bootstrapping Evolvability for Inter-domain Routing with D-BGP}}. In \bibinfo{booktitle}{\emph{Proceedings of SIGCOMM}}. ACM.
\newblock


\bibitem[Savage et~al\mbox{.}(1999)]%
        {detour}
\bibfield{author}{\bibinfo{person}{S. Savage}, \bibinfo{person}{T. Anderson}, \bibinfo{person}{A. Aggarwal}, \bibinfo{person}{D. Becker}, \bibinfo{person}{N. Cardwell}, \bibinfo{person}{A. Collins}, \bibinfo{person}{E. Hoffman}, \bibinfo{person}{J. Snell}, \bibinfo{person}{A. Vahdat}, \bibinfo{person}{G. Voelker}, {and} \bibinfo{person}{J. Zahorjan}.} \bibinfo{year}{1999}\natexlab{}.
\newblock \showarticletitle{{Detour: Informed Internet Routing and Transport}}.
\newblock \bibinfo{journal}{\emph{IEEE Micro}} (\bibinfo{year}{1999}).
\newblock


\bibitem[Schlinker et~al\mbox{.}(2019)]%
        {schlinker19peering}
\bibfield{author}{\bibinfo{person}{Brandon Schlinker}, \bibinfo{person}{Todd Arnold}, \bibinfo{person}{Italo Cunha}, {and} \bibinfo{person}{Ethan Katz-Bassett}.} \bibinfo{year}{2019}\natexlab{}.
\newblock \showarticletitle{{PEERING: Virtualizing BGP at the Edge for Research}}. In \bibinfo{booktitle}{\emph{Proceedings of CoNEXT}}. ACM.
\newblock


\bibitem[Schlinker et~al\mbox{.}(2017)]%
        {schlinker2017edge}
\bibfield{author}{\bibinfo{person}{Brandon Schlinker}, \bibinfo{person}{Hyojeong Kim}, \bibinfo{person}{Timothy Cui}, \bibinfo{person}{Ethan Katz-Bassett}, \bibinfo{person}{Harsha~V. Madhyastha}, \bibinfo{person}{Italo Cunha}, \bibinfo{person}{James Quinn}, \bibinfo{person}{Saif Hasan}, \bibinfo{person}{Petr Lapukhov}, {and} \bibinfo{person}{Hongyi Zeng}.} \bibinfo{year}{2017}\natexlab{}.
\newblock \showarticletitle{{Engineering Egress with Edge Fabric: Steering Oceans of Content to the World}}. In \bibinfo{booktitle}{\emph{Proceedings of SIGCOMM}}. ACM.
\newblock


\bibitem[Shao and Gao(2021)]%
        {gao2021policy}
\bibfield{author}{\bibinfo{person}{Xiaozhe Shao} {and} \bibinfo{person}{Lixin Gao}.} \bibinfo{year}{2021}\natexlab{}.
\newblock \showarticletitle{{Policy-Rich Interdomain Routing with Local Coordination}}.
\newblock \bibinfo{journal}{\emph{Computer Networks}} (\bibinfo{year}{2021}).
\newblock


\bibitem[Spring et~al\mbox{.}(2003)]%
        {SIGCOMM03-PathInflation}
\bibfield{author}{\bibinfo{person}{Neil Spring}, \bibinfo{person}{Ratul Mahajan}, {and} \bibinfo{person}{Thomas Anderson}.} \bibinfo{year}{2003}\natexlab{}.
\newblock \showarticletitle{{The Causes of Path Inflation}}. In \bibinfo{booktitle}{\emph{Proceedings of SIGCOMM}}. ACM, \bibinfo{pages}{113–124}.
\newblock


\bibitem[Staff(2015)]%
        {RIPE}
\bibfield{author}{\bibinfo{person}{RIPE~NCC Staff}.} \bibinfo{year}{2015}\natexlab{}.
\newblock \showarticletitle{{RIPE Atlas: A Global Internet Measurement Network}}.
\newblock \bibinfo{journal}{\emph{Internet Protocol Journal}}  \bibinfo{volume}{18} (\bibinfo{year}{2015}).
\newblock


\bibitem[Villamizar et~al\mbox{.}(1998)]%
        {RFC2439-RouteDamping}
\bibfield{author}{\bibinfo{person}{Curtis Villamizar}, \bibinfo{person}{Ravi Chandra}, {and} \bibinfo{person}{Ramesh Govindan}.} \bibinfo{year}{1998}\natexlab{}.
\newblock \bibinfo{booktitle}{\emph{{RFC 2439: BGP Route Flap Damping}}}.
\newblock \bibinfo{type}{Internet Requests for Comments}. \bibinfo{institution}{Internet Engineering Task Force (IETF)}.
\newblock


\bibitem[Walton et~al\mbox{.}(2016)]%
        {RFC7911-BGPAddPath}
\bibfield{author}{\bibinfo{person}{Daniel Walton}, \bibinfo{person}{Alvaro Retana}, \bibinfo{person}{Enke Chen}, {and} \bibinfo{person}{John Scudder}.} \bibinfo{year}{2016}\natexlab{}.
\newblock \bibinfo{booktitle}{\emph{{RFC 7911: Advertisement of Multiple Paths in BGP}}}.
\newblock \bibinfo{type}{Internet Requests for Comments}. \bibinfo{institution}{Internet Engineering Task Force (IETF)}.
\newblock


\bibitem[Yang et~al\mbox{.}(2007)]%
        {nira}
\bibfield{author}{\bibinfo{person}{Xiaowei Yang}, \bibinfo{person}{David Clark}, {and} \bibinfo{person}{Arthur~W. Berger}.} \bibinfo{year}{2007}\natexlab{}.
\newblock \showarticletitle{NIRA: A New Inter-Domain Routing Architecture}.
\newblock \bibinfo{journal}{\emph{IEEE/ACM Transactions on Networking}} \bibinfo{volume}{15}, \bibinfo{number}{4} (\bibinfo{year}{2007}), \bibinfo{pages}{775--788}.
\newblock


\bibitem[Yap et~al\mbox{.}(2017)]%
        {espresso}
\bibfield{author}{\bibinfo{person}{Kok-Kiong Yap}, \bibinfo{person}{Murtaza Motiwala}, \bibinfo{person}{Jeremy Rahe}, \bibinfo{person}{Steve Padgett}, \bibinfo{person}{Matthew Holliman}, \bibinfo{person}{Gary Baldus}, \bibinfo{person}{Marcus Hines}, \bibinfo{person}{Taeeun Kim}, \bibinfo{person}{Ashok Narayanan}, \bibinfo{person}{Ankur Jain}, {et~al\mbox{.}}} \bibinfo{year}{2017}\natexlab{}.
\newblock \showarticletitle{{Taking the Edge off with Espresso: Scale, Reliability and Programmability for Global Internet Peering}}. In \bibinfo{booktitle}{\emph{Proceedings of SIGCOMM}}. ACM, \bibinfo{pages}{432--445}.
\newblock


\bibitem[Zhou et~al\mbox{.}(2023)]%
        {SIGCOMM23-RegionalCDN}
\bibfield{author}{\bibinfo{person}{Minyuan Zhou}, \bibinfo{person}{Xiao Zhang}, \bibinfo{person}{Shuai Hao}, \bibinfo{person}{Xiaowei Yang}, \bibinfo{person}{Jiaqi Zheng}, \bibinfo{person}{Guihai Chen}, {and} \bibinfo{person}{Wanchun Dou}.} \bibinfo{year}{2023}\natexlab{}.
\newblock \showarticletitle{Regional IP Anycast: Deployments, Performance, and Potentials}. In \bibinfo{booktitle}{\emph{Proceedings of SIGCOMM}}. ACM, \bibinfo{pages}{917--931}.
\newblock


\end{thebibliography}

\end{document}